\documentclass[12pt,preprint]{aastex}

\begin{document}

\slugcomment{Accepted for the Astrophysical Journal}

\title{Energetic Protons, Radionuclides and Magnetic Activity in
  Protostellar Disks}

\author{N.~J.~Turner} \affil{Jet Propulsion Laboratory, California
  Institute of Technology, Pasadena, California 91109;
  neal.turner@jpl.nasa.gov}

\and \author{J.~F.~Drake} \affil{Department of Physics, Institute for
  Physical Science and Technology, University of Maryland, College
  Park, Maryland 20742}

\begin{abstract}
  We calculate the location of the magnetically-inactive dead zone in
  the minimum-mass protosolar disk, under ionization scenarios
  including stellar X-rays, long- or short-lived radionuclide decay,
  and energetic protons arriving from the general interstellar medium,
  from a nearby supernova explosion, from the disk corona, or from the
  corona of the young star.  The disk contains a dead zone in all
  scenarios except those with small dust grains removed and a fraction
  of the short-lived radionuclides remaining in the gas.  All the
  cases without exception have an ``undead zone'' where intermediate
  resistivities prevent magneto-rotational turbulence while allowing
  shear-generated large-scale magnetic fields.  The mass column in the
  undead zone is typically greater than the column in the turbulent
  surface layers.  The results support the idea that the dead and
  undead zones are robust consequences of cold, dusty gas with mass
  columns exceeding 1000~g~cm$^{-2}$.
\end{abstract}


\section{INTRODUCTION\label{sec:intro}}

Angular momentum transport is a key to the evolution of protostellar
disks and the origins of the planets, as it governs the flow of raw
materials toward the star \citep{b95}.  Transport processes likely to
be important include the turbulence resulting from the
magneto-rotational instability \citep[MRI;][]{bh91,bh98} and the
stresses due to large-scale magnetic fields driving an outflow
\citep{bp82,wk93} or shearing within the disk \citep{ts08}.

All of the magnetic angular momentum transport processes require the
gas is sufficiently ionized to couple to the fields.  Collisional
ionization leads to good coupling only at the temperatures above
1000~K found very close to the star \citep{pm65,u83,nu88}.  In the
overwhelming majority of the disk volume, the main ionization
processes are non-thermal.  Among the important non-thermal processes
are the X-rays emitted by the young star \citep{gn97}, the cosmic rays
arriving from interstellar space \citep{u83,un88}, and the decay of
radionuclides within the gas \citep{un81}.  The basic difficulties in
reaching high enough levels of ionization through these non-thermal
processes are the large column of absorbing material and the high rate
of recombination on the surfaces of dust particles.  The X-rays and
cosmic rays generally penetrate and ionize only the surface layers of
the disk, leaving a region near the midplane where magnetic activity
is suppressed \citep{g96,sm00,in06a}.

The absorption of X-rays and cosmic rays in the atmosphere means that
the disk can be divided into three zones.  In the upper layers, and in
outer annuli where the mass column is low enough for the ionizing
radiation to reach the midplane, the magnetic fields couple thoroughly
to the gas, and magnetic forces drive turbulence through the MRI.  In
a ``dead zone'' near the midplane, the fields decouple from the gas,
and magnetic forces are largely irrelevant.  At resistivities between
these extremes, magnetic fields decouple over scales comparable to the
disk thickness, shutting off the MRI, but remain coupled over the disk
radius.  The radial gradient in orbital frequency can then shear out
any weak radial magnetic field to generate toroidal fields, enabling
angular momentum transport to continue \citep{ts08}.  The turbulent
surface layers are thus separated from the midplane dead zone by an
``undead zone'' that becomes magnetically active when supplied with
radial fields.

In this paper we explore whether three less well-studied additional
sources of ionization can reduce the sizes of the dead and undead
zones or make protostellar disks magnetically active throughout.  The
three ionization sources are high-energy particles from (1) a nearby
supernova explosion \citep{fa06}, (2) the corona of the protostellar
disk itself, and (3) the young star.  We estimate the ionization rates
to order of magnitude (\S\ref{sec:ionization}) and calculate the
resulting resistivities (\S\ref{sec:chemistry}) in the minimum-mass
model of the protosolar disk (\S\ref{sec:disk}), finding the undead
and dead zones shown in \S\ref{sec:deadzone}.  A summary and
conclusions are in \S\ref{sec:conclusions}.

\section{IONIZATION\label{sec:ionization}}

The base ionization rate in our calculations is set by the X-rays
emitted from the vicinity of the young star (\S\ref{sec:xrays}).  Some
of the models include also the decay of radioactive isotopes within
the disk (\S\ref{sec:radionuclides}).  Our purpose is to explore the
additional effects of energetic protons from three sources: the
interstellar medium (\S\ref{sec:cr}), the disk corona
(\S\ref{sec:diskcorona}), and the stellar corona
(\S\ref{sec:stellarcorona}).  With the aim of determining whether the
dead zone can be removed under favorable circumstances, we generally
err on the high side in estimating the proton fluxes.  The rates of
ionization from the five processes are compared in
\S\ref{sec:compared}.

\subsection{Stellar X-Rays\label{sec:xrays}}

The X-ray ionization is included because almost all lightly-obscured
young solar-mass stars in the Orion nebula show X-ray emission
\citep{gf00,pk05}.  The measured spread in luminosity is several
decades, with a median around $2\times 10^{30}$~erg~s$^{-1}$.
Evidence that the X-rays strike the disks comes from the iron
K$\alpha$ fluorescent emission detected in some pre-main-sequence
stars having near-infrared excesses \citep{ik01,fm05,tf05,gf07}.  Gas
at distance $r$ directly facing the star is ionized at a rate $\zeta_X
= \sigma F_X/E_i \approx 5\times 10^{-11}\ {\rm s}^{-1}\ (r/{\rm
  AU})^{-2}$, using a flux $F_X$ corresponding to the median X-ray
luminosity, an energy requirement per ionization $E_i=37$~eV, and an
absorption cross-section $\sigma=4.16\times 10^{-24}$~cm$^2$ at photon
energy 5~keV and Solar composition \citep{gn97}.  Most of the disk
however is ionized at much lower rates, because the photons enter at
grazing angles and only the most energetic and least numerous reach
the deep interior.  Detailed Monte Carlo radiative transfer results
for a 5~keV thermal source spectrum \citep{ig99}, fitted assuming an
inverse square falloff with the radius, yield an ionization rate
\begin{equation}
  \zeta_X=2.6\times 10^{-15}\ {\rm s}^{-1}\ (r/{\rm AU})^{-2}
  \ [\exp(-\Sigma_a/\Sigma_X) + \exp(-\Sigma_b/\Sigma_X)]
\end{equation}
at columns greater than 1~g~cm$^{-2}$, where
$\Sigma_X=8.0$~g~cm$^{-2}$ is the X-ray absorption depth and
$\Sigma_a$ and $\Sigma_b$ are the mass columns lying vertically above
and below the point of interest.

\subsection{Radionuclides\label{sec:radionuclides}}

The decay of long-lived radionuclides, primarily $^{40}$K (half-life
1.25~Gyr), gives an ionization rate $\zeta_{LR}=6.9\times
10^{-23}$~s$^{-1}$ at protosolar elemental abundances \citep{un81}.
Short-lived radionuclides, primarily $^{26}$Al (half-life 0.717~Myr),
yield an ionization rate $\zeta_{SR}=3.7\times 10^{-19}$~s$^{-1}$
\citep{s92}.  The results of our calculations change little if we use
instead the rates advocated by \cite{un09}, which differ from these by
between 9\% and 55\%.  The radionuclide ionization rates will be
substantially lower if the dust particles containing most of the heavy
elements are absent.  In some cases without grains we include no
radionuclide ionization, while in others the rate is reduced
$10\,000$-fold from the dusty calculations to account for the
possibility that a small amount of the radioactive element occurs in
the gas phase.

\subsection{Interstellar Cosmic Rays\label{sec:cr}}

The first of three energetic proton sources we consider is the
interstellar cosmic rays.  Particles with energies above 0.1~GeV
provide most of the ionizing effect at high mass columns \citep{un81}.
Shielding by the contemporary Solar wind reduces the flux of
interstellar cosmic rays reaching the top of the Earth's atmosphere in
the energy range up to a few GeV \citep{st68}.  Estimates of the
cosmic ray energy spectrum outside the heliosphere, reviewed by
\cite{if09}, scatter around the levels inferred by \cite{ch72} that we
use here.  The intensity $0.94(E_0+E)^{-2.6}$ cm$^{-2}$ s$^{-1}$
sr$^{-1}$ GeV$^{-1}$ falls off with energy $E$ above the proton rest
energy $E_0=0.938$~GeV.  The particles with $E>0.1$~GeV deliver an
energy flux of approximately 0.01 erg cm$^{-2}$ s$^{-1}$, producing an
ionization rate
\begin{equation}\label{eqn:zetacr}
\zeta_{CR}=5\times 10^{-18}\ {\rm s}^{-1}
\left\{
 \exp\left({-\Sigma_a\over \Sigma_{CR}}\right)
 \left[1+\left(\Sigma_a\over\Sigma_{CR}\right)^{3\over 4}\right]^{-{4\over 3}} +
 \exp\left({-\Sigma_b\over \Sigma_{CR}}\right)
 \left[1+\left(\Sigma_b\over\Sigma_{CR}\right)^{3\over 4}\right]^{-{4\over 3}}
\right\}
\end{equation}
under isotropic illumination of the disk surface \citep{un09}.  The
cosmic ray absorption depth $\Sigma_{CR}=96$~g~cm$^{-2}$ \citep{un81}.
Lacking good information about the degree to which the protosolar wind
screened the inner protosolar disk from interstellar cosmic rays, and
seeking a lower bound on the size of the undead and dead zones, we
assume that the cosmic rays above 0.1~GeV reach the disk surface with
the interstellar flux and energy spectrum.

A supernova explosion occurring within 10~pc of the young star could
increase the cosmic ray energy density by about three orders of
magnitude over the general interstellar value, based on the
energetic-particle populations inferred from gamma-ray observations of
supernova remnants \citep{fa06}.  We therefore consider below the
effects of a thousand-fold increase in the cosmic ray ionization rate.

\subsection{Disk Corona\label{sec:diskcorona}}

The second energetic proton source we consider is the atmosphere of
the accretion disk itself.  Forbidden optical line emission and
central Balmer absorptions in the spectra of T~Tauri stars indicate
the presence of low-density overlying gas with temperatures in the
range $5\,000$ to $10\,000$~K \citep{k97} similar to the Solar
chromosphere.  Models of the heating and cooling balance show that
such temperatures can be produced by the stellar ultraviolet and X-ray
irradiation \citep{kd04,gn04,na07}.  Furthermore, many T~Tauri disks
produce winds with speeds of 10 to 100~km~s$^{-1}$
\citep{he95,tb01,ph02,af07}, typically surrounding a faster central
jet.  Since magnetic forces likely drive the winds, we assume below
that the Alfv\'en speeds approach the escape speed in the extended
atmospheres of the disks.

Under similar conditions in the Solar chromosphere and corona, fast
magnetic reconnection and shocks lead to the acceleration of protons
to high energies, as reviewed by \cite{mc97}.  Fast reconnection
requires plasma in the Hall regime, so that the whistler or kinetic
Alfv\'en driving waves are not dissipated \citep{cs05}.  The
electron-ion collision frequency $\nu_{ei}$ must be less than the
electron cyclotron frequency $\Omega_{ce}$.  In the atmosphere of the
disk around a young Solar-mass star, the ratio of the two frequencies
is
\begin{equation}\label{eqn:reconn}
{\nu_{ei}\over\Omega_{ce}} \approx 0.01
\left(n\over 10^{10}\,{\rm cm}^{-3}\right)
\left(T\over 10^4\,{\rm K}\right)^{-3/2}
\left(r\over 1\,{\rm AU}\right)^{1/2},
\end{equation}
using expressions from \cite{mk73} with the Coulomb logarithm set
to~10.  Fast reconnection can thus occur in the tenuous disk
atmosphere.  Reconnection in the dense, cold disk interior presumably
is due to the slower Sweet-Parker mechanism.

Reconnection in the Solar corona produces a mix of thermal and
non-thermal particles.  The temperature of the thermal component is
such that the downstream gas pressure is approximately equal to the
upstream magnetic pressure.  Similar conversion of magnetic energy to
heat in the disk corona yields the virial temperature $10\,{\rm eV}\
(r/{\rm AU})^{-1}$.  Reconnection can thus plausibly contribute to the
heating and stirring indicated by the central Balmer absorptions
\citep{k97} and by the Doppler-broadened emission line of ionized neon
at 12.8~$\mu$m in TW~Hydrae \citep{hn07}.  The thermal particles do
not affect the dead zone as they have little penetrating power.

The maximum energy of the non-thermal particles is approximately equal
to the product of the particle charge $q$, the electric field
strength, and the length over which the particles are accelerated.
The electric field $|{\bf v \times B}|/c$ is proportional to the
reconnection speed $|{\bf v}|$, which we set equal to the Alfv\'en
speed $v_A$.  The lengths of magnetic loops in the disk corona depend
on the rate of reconnection relative to shear \citep{ug08}.  When
shear dominates, loops may reach sizes comparable to the radius $r$,
which we choose as the acceleration distance.  The maximum energy of
the particles produced during reconnection
\begin{equation}
  E_{rec}=q \left({v_A B \over c}\right) r
  \approx 100\,{\rm GeV}\ \left(n\over 10^{10}{\rm cm}^{-3}\right)^{1/2},
\end{equation}
so the high-energy tail extends into the regime of penetrating cosmic
rays above $0.1$~GeV.

The luminosity of the particles accelerated by reconnection in the
disk corona is bounded above by the rate of gravitational energy
release due to accretion.  The accretion energy per unit area and unit
time leaving each side of the disk around a young solar analog T~Tauri
star is about $3\times 10^5 ({\dot M}/10^{-8} M_\odot$yr$^{-1})
(r/{\rm AU})^{-3}$ times the energy flux in ordinary interstellar
cosmic rays, where $\dot M$ is the mass flow rate.  Much of the
accretion flux takes the form of thermal radiation from the disk
photosphere, while some fraction is in magnetic fields that rise into
the corona before dissipating.  Results from shearing-box MHD
calculations \citep{ms00} suggest the fraction is of order 10\%.  Some
part of the coronal magnetic energy is converted through reconnection
and shocks into the kinetic energy of non-thermal protons.  We assume
the proportion is similar to the contemporary Sun, where magnetic
energy is dissipated at approximately $10^{27}$ erg~s$^{-1}$, judging
from the mean X-ray luminosity.  The Sun and solar wind produce a flux
at 1~AU of about one proton with energy exceeding 0.1~GeV, per square
centimeter per second, comparable to the interstellar cosmic ray flux
outside the heliosphere, according to data from the GOES satellite
energetic particle sensors \citep{mc07} and to analyses of tracks
produced in lunar rocks over the past few million years \citep{rm91}.
The corresponding solar luminosity in particles above 0.1~GeV is
around $10^{24}$ erg~s$^{-1}$, or 0.1\% of the rate of dissipation of
magnetic energy.  We thus infer that the overall fraction of the disk
accretion energy that rises into the corona in the form of magnetic
fields and is converted to energetic protons is $10^{-4}$.  Perhaps
half the particles travel downward and strike the disk surface,
yielding an ionisation rate $\zeta_{DP}$ that is $10(r/{\rm AU})^{-3}$
times the demodulated interstellar cosmic ray value.

\subsection{Stellar Corona\label{sec:stellarcorona}}

The third source of energetic protons we consider is the corona and
inner wind of the young Sun.  The contemporary Sun accelerates ions to
cosmic ray energies.  The reconnection of opposing magnetic fields
with Alfv\'en speed $v_A$ produces ions of mass $m$ with energies of
order $m v_A^2$, according to observations near 1~AU in the Solar wind
\citep{gs05} and to particle-in-cell models \citep{kw07,ds09}.
Flaring regions in the Solar corona have densities of $10^9$ to
$10^{10}$~cm$^{-3}$ \citep{bb83,ks08} and fields around 50~G, yielding
protons with energies in the tens of keV range.  How these seed
particles reach much higher energies remains unclear.  Ideas include
repeated shock crossings \citep[as reviewed by][]{md01} and magnetic
reconnection \citep{lo09,do09}.

The Sun produced more energetic particles when young: the X-ray flares
observed in young stars, scaled by the solar relationship between
X-ray luminosity and proton emission, indicate a $10^5$-fold
enhancement in energetic protons compared to contemporary solar levels
\citep{fg02}.  Such an enhancement appears to be consistent with a
spallation origin for $^{26}$Al and other short-lived radionuclides
whose decay products occur in primitive chondritic meteorites
\citep{gs06}.  In addition, the magnetic fields of young Solar-mass
stars are strong enough for reconnection to produce large numbers of
seed particles with energies above an MeV.  T~Tauri stars have
kilogauss surface-averaged magnetic fields, three orders of magnitude
stronger than the modern Sun \citep{jv99,j07}, while the densities of
gas on coronal magnetic loops, inferred from the emitted X-rays, are
around $10^{10}$ cm$^{-3}$ \citep{ff05,jc06}.  Absorption of the
X-rays by foreground coronal material indicates columns also
consistent with such densities \citep{rs07}.  The characteristic
energy of the seed protons produced in reconnection under these
conditions is 10~MeV if the fields are uniformly-distributed, or
larger if as is likely the fields are patchy across the face of the
star.

Lacking better knowledge of the arrangement of the magnetic fields
near the star, we cannot say whether the stellar coronal protons are
channeled back into the star, out along open field lines into a wind,
or toward the disk which may reach or penetrate the boundary of the
stellar magnetosphere.  Since our objective is a minimum size for the
dead zone, we shall assume that the protons propagate in straight
lines like the X-rays, reaching all parts of the disk surface
unimpeded.

One must also bear in mind that most of the protons are produced in
flares lasting only hours or days.  The flare particles will reduce
the size of the dead zone in the disk if the ionization created in one
flare persists till the next.  Calculations of the ionization due to
X-ray flares suggest that the recombination time is less than the
interval between flares unless the abundance of the dust in the disk
is greatly reduced \citep{in06c}.  A more conservative estimate of the
contribution of the stellar energetic particles comes from scaling the
modern Sun up to the X-ray luminosity corresponding to $\zeta_X$.  The
effective enhancement in the energetic protons is then only about
3.3~decades and the rate of ionization is comparable to that resulting
from the stellar X-rays.

Nevertheless, in the interest of obtaining a minimum dead zone size,
we treat the $10^5$-fold enhancement in the energetic protons as
time-steady.  The flux falls off roughly as the inverse square of the
distance, at distances much greater than the size of the stellar
magnetosphere.  Projection effects reduce the flux per unit disk area
by about a factor ten, since the disk surface is inclined about
0.1~radian from the line of sight to the star.  We take a surface
ionisation rate $\zeta_{SP}$ of $10^4(r/{\rm AU})^{-2}$ times the
interstellar cosmic ray value.  The particles from the coronae of the
disk and star are assumed to decline with column in the same way as
the interstellar cosmic rays, following eq.~\ref{eqn:zetacr}.

\subsection{Ionization Processes Compared\label{sec:compared}}

\begin{figure}[tb!]
  \epsscale{0.6}
  \plotone{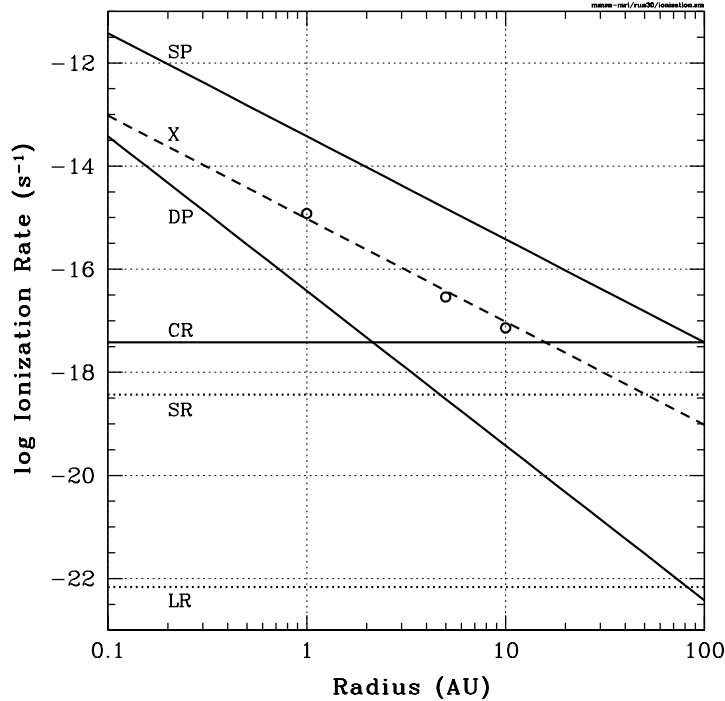}

  \figcaption{\sf Ionization rates at column 8~g~cm$^{-2}$ versus the
    distance from the star.  The dashed line shows the contribution of
    stellar X-rays, and the dotted lines the decay of short-lived
    (upper) and long-lived (lower) radionuclides.  Among the three
    solid lines, the uppermost is for stellar energetic particles, the
    steepest shows the disk coronal particles, and the horizontal
    solid line is for interstellar cosmic rays.  Large surface density
    is assumed and only radiation striking the top of the disk is
    included on this diagram.  Circles indicate the results of the
    Monte Carlo X-ray transfer calculations by \cite{ig99}, scaled to
    a stellar X-ray luminosity $2\times 10^{30}$ erg~s$^{-1}$.  The
    disk coronal particle estimate corresponds to a fraction $10^{-4}$
    of the energy released by mass accretion at rate $10^{-8}$
    M$_\odot$~yr$^{-1}$.
    \label{fig:compared}}
\end{figure}

The relative strengths of the different ionization processes are shown
in figure~\ref{fig:compared} where the ionization rates at a column of
8~g~cm$^{-2}$ are plotted versus radius.  Compared with the stellar
X-rays, the stellar energetic protons are 40~times stronger and are
the strongest process plotted.  The disk coronal energetic protons are
a few times weaker than the X-rays at 0.1~AU and fall off faster with
distance.  The interstellar cosmic rays are distance-independent and
are stronger than the X-rays outside 16~AU.  If enhanced a thousand
times due to a nearby supernova, interstellar cosmic rays are stronger
than the stellar X-rays outside 0.5~AU and are the strongest of all
the processes outside 3~AU.

\section{RESISTIVITY\label{sec:chemistry}}

Recombination in the gas phase and on grain surfaces balances the
ionization processes in local chemical equilibrium.  We find the
balance by solving the reduced chemical reaction network discussed by
\cite{in06a}.  The ionization produces molecular ions which are
destroyed through dissociative recombination, or by charge exchange
with metal atoms which then recombine radiatively.  Magnesium is
chosen as a representative gas-phase metal because of its combination
of high abundance and low desorption temperature.  The amount
initially in the gas is chosen to be 1\% of the Solar magnesium
abundance.  Recombination also occurs on grains which can collide with
electrons, ions and other grains.  The grains are 1~$\mu$m in radius
with an internal density 3 g~cm$^{-3}$ and a mass fraction 1\%
relative to the gas.  In addition, molecules and atoms can become
adsorbed on grain surfaces, reducing the gas phase abundances.  The
stiff set of ODEs representing the chemical kinetic equations is
integrated to equilibrium using semi-implicit extrapolation
\citep{pt92}.  The resulting electrical resistivity $\eta =
234\sqrt{T}/x_e$~cm$^2$~s$^{-1}$ varies inversely with the electron
fraction $x_e=n_e/n_n$, where $n_e$ is the electron number density and
$n_n$ the total number density of neutrals \citep{bb94}.

\section{DISK MODEL\label{sec:disk}}

We use the minimum-mass model of the protosolar disk \citep{hn85},
with surface mass density $\Sigma=1700$~g~cm$^{-2}$ $(r/{\rm
  AU})^{-3/2}$ and temperature $T=280$~K $(r/{\rm AU})^{-1/2}$ at
distance $r$ measured in astronomical units from a Solar-mass star.
Vertical hydrostatic balance yields a volume density
$\rho=(\Sigma/\sqrt{2\pi}H) \exp(-z^2/2 H^2)$ declining with height
$z$ as a Gaussian, with a characteristic scale $H=c_s/\Omega$ that is
the ratio of the isothermal sound speed $c_s$ to the orbital frequency
$\Omega$.

\section{DEAD ZONE\label{sec:deadzone}}

Magnetic stresses extract angular momentum from the disk only if the
gas is sufficiently ionized to couple to the magnetic fields.  Upper
bounds on the allowed Ohmic resistivity $\eta$ come from considering
the dissipation scale.  The minimum time for regenerating the field is
the orbital period, while in the case of global magnetic fields, the
characteristic size of the variations in the field is the distance to
the star.  The orbital shear acting on a weak radial seed field can
produce large-scale toroidal fields faster than the fields diffuse
away if
\begin{equation}\label{eqn:luvk10}
{v_K^2 \over \eta\Omega} > 10
\end{equation}
where $v_K$ is the Keplerian orbital speed, $\Omega$ the orbital
frequency and power-law radial variation is assumed for the magnetic
fields \citep{ts08}.  Using a similar approach, MRI turbulence
regenerates tangled magnetic fields if the resistivity is unable to
erase magnetic variations over the MRI fastest-growing vertical
wavelength $2\pi v_{Az}/\Omega$, where $v_{Az}$ is the vertical
component of the Alfv\'en speed.  The criterion for MRI turbulence
\begin{equation}\label{eqn:luva1}
{v_{Az}^2 \over \eta\Omega} > 1
\end{equation}
depends on the volume-averaged squared vertical Alfven speed, whether
the background magnetic field is vertical, toroidal, or zero
\citep{ss02b}.  Of eqs.~\ref{eqn:luvk10} and~\ref{eqn:luva1}, the
turbulence condition is the more restrictive since the MRI wavelength
typically is much less than the radius.  The two criteria divide the
disk into three unequal zones.  In the active zone both conditions
hold and strong magnetic activity is expected.  In the undead zone the
shear condition eq.~\ref{eqn:luvk10} holds and the turbulence
condition eq.~\ref{eqn:luva1} does not.  Magnetic fields can then be
generated by shear but not by MRI turbulence.  In the dead zone,
neither condition is satisfied and the gas is almost completely
decoupled from the magnetic fields.

We determine the spatial extent of the magnetic activity in the
minimum-mass protosolar disk by finding the locations where
eqs.~\ref{eqn:luvk10} and~\ref{eqn:luva1} are marginally satisfied.
The ionization-recombination reaction network is solved to determine
the resistivity on a grid of $60\times 250$ points spaced
logarithmically in radius and linearly in $z/H$, from $r=0.1$~AU to
100~AU and from $z=0$ to $5 H$.  The Alfv\'en speed in
eq.~\ref{eqn:luva1} is calculated assuming the vertical magnetic
fields at each radius have a pressure 1000~times less than the
midplane gas pressure.  Similar pressure ratios are found in ideal-MHD
stratified shearing-box calculations \citep{ms00}.

\begin{figure}[tb!]
  \epsscale{0.6}
  \plotone{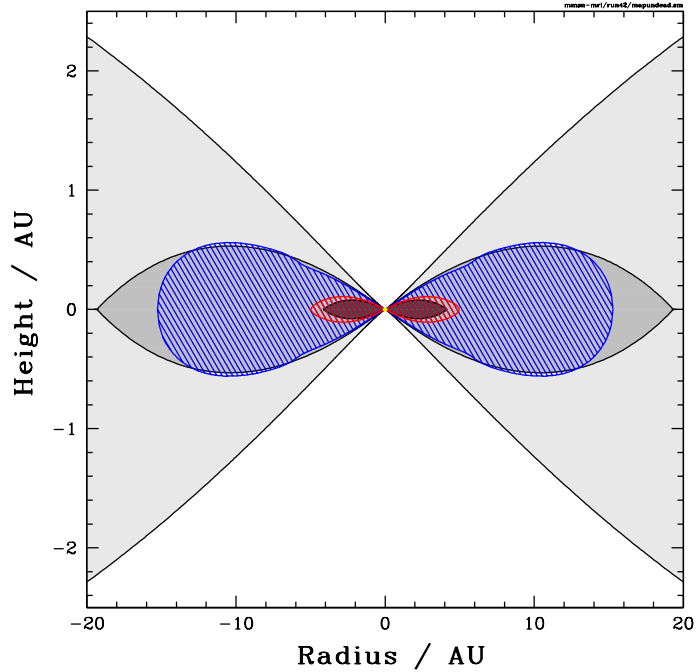}

  \figcaption{\sf Dead zone (red) and undead zone (blue) in the
    minimum-mass protosolar disk with well-mixed 1-$\mu$m dust grains
    at a mass fraction 1\%.  The ionization is due to stellar X-rays
    and long-lived radionuclides, and the vertical component of the
    magnetic field has pressure 0.1\% of the midplane gas pressure.
    The vertical mass column exceeds 1, 10 and 100~g~cm$^{-2}$ in the
    light, medium and dark gray shaded regions, respectively.  Note
    that the vertical scale is expanded by a factor four.
    \label{fig:map}}
\end{figure}

We first compute the sizes of the undead and dead zones in a fiducial
model disk including ionization by stellar X-rays and long-lived
radionuclides, and recombination on grains.  For this calculation
only, the chemical network is solved on a finer grid of $1000\times
1000$ points to more precisely determine the shape of the dead zone
boundary.  The results appear in figure~\ref{fig:map}.  MRI turbulence
reaches the midplane only outside 15~AU, while large-scale magnetic
fields couple to the orbital shear at the midplane outside 5~AU.  The
boundary of the turbulent region approximately follows the contour of
vertical column 10~g~cm$^{-2}$, close to the X-ray absorption depth of
8~g~cm$^{-2}$.  Beyond 12~AU, turbulence occurs at greater columns
because the lower densities mean slower recombination.

\begin{figure}[tb!]
  \epsscale{0.6}
  \plotone{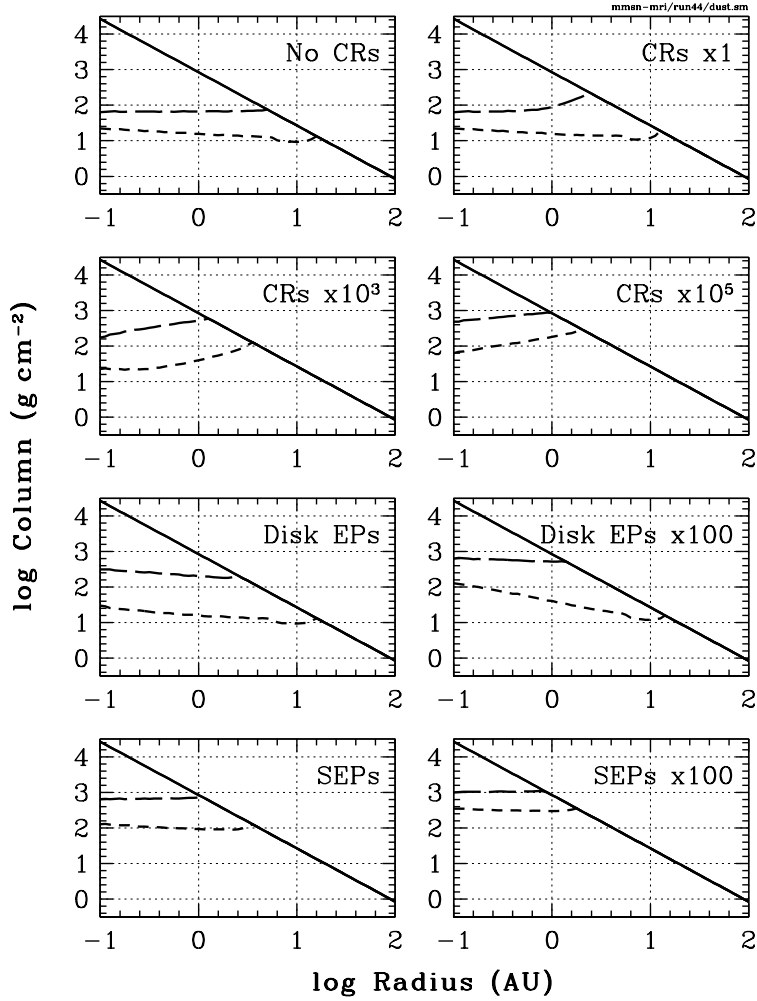}

  \figcaption{\sf Column depth to the top of the undead zone (short
    dashes) and dead zone (long dashes) in the minimum-mass protosolar
    disk with well-mixed 1~$\mu$m dust grains at a mass fraction 1\%.
    Solid lines show the midplane column.  All eight ionization
    scenarios include stellar X-rays and long-lived radionuclides.
    The scenarios are (left to right and top to bottom): (1) no
    additional ionization, (2) with interstellar cosmic rays, (3) with
    cosmic ray flux enhanced 1000~times due to a nearby supernova, (4)
    with cosmic ray flux enhanced a further two orders of magnitude,
    (5) with energetic particles from the disk corona, (6) with the
    disk particles enhanced 100~times, (7) with energetic particles
    from the young Sun, and (8) with the protosolar particles enhanced
    100~times.
    \label{fig:dust}}
\end{figure}

Results for eight different external ionization scenarios are shown in
figure~\ref{fig:dust}.  All the scenarios include stellar X-ray
ionization.  The first has no other external sources of ionization,
and corresponds to figure~\ref{fig:map}.  The second includes ordinary
interstellar cosmic rays.  The third and fourth have the cosmic rays
enhanced by factors $10^3$ and $10^5$.  The fifth and sixth instead
have energetic particles from the disk corona, one with the scaling
indicated in section~\ref{sec:diskcorona} and the other with an
ionization rate 100~times greater, corresponding to a higher mass
accretion rate of $10^{-6}$ M$_\odot$~yr$^{-1}$.  The seventh and
eighth instead have protosolar energetic particles, one with the rate
from section~\ref{sec:stellarcorona} and the other with a rate
100~times greater.  All the calculations in figure~\ref{fig:dust} also
include ionization by long-lived radionuclides, and recombination on
grains.

\begin{figure}[tb!]
  \epsscale{0.6}
  \plotone{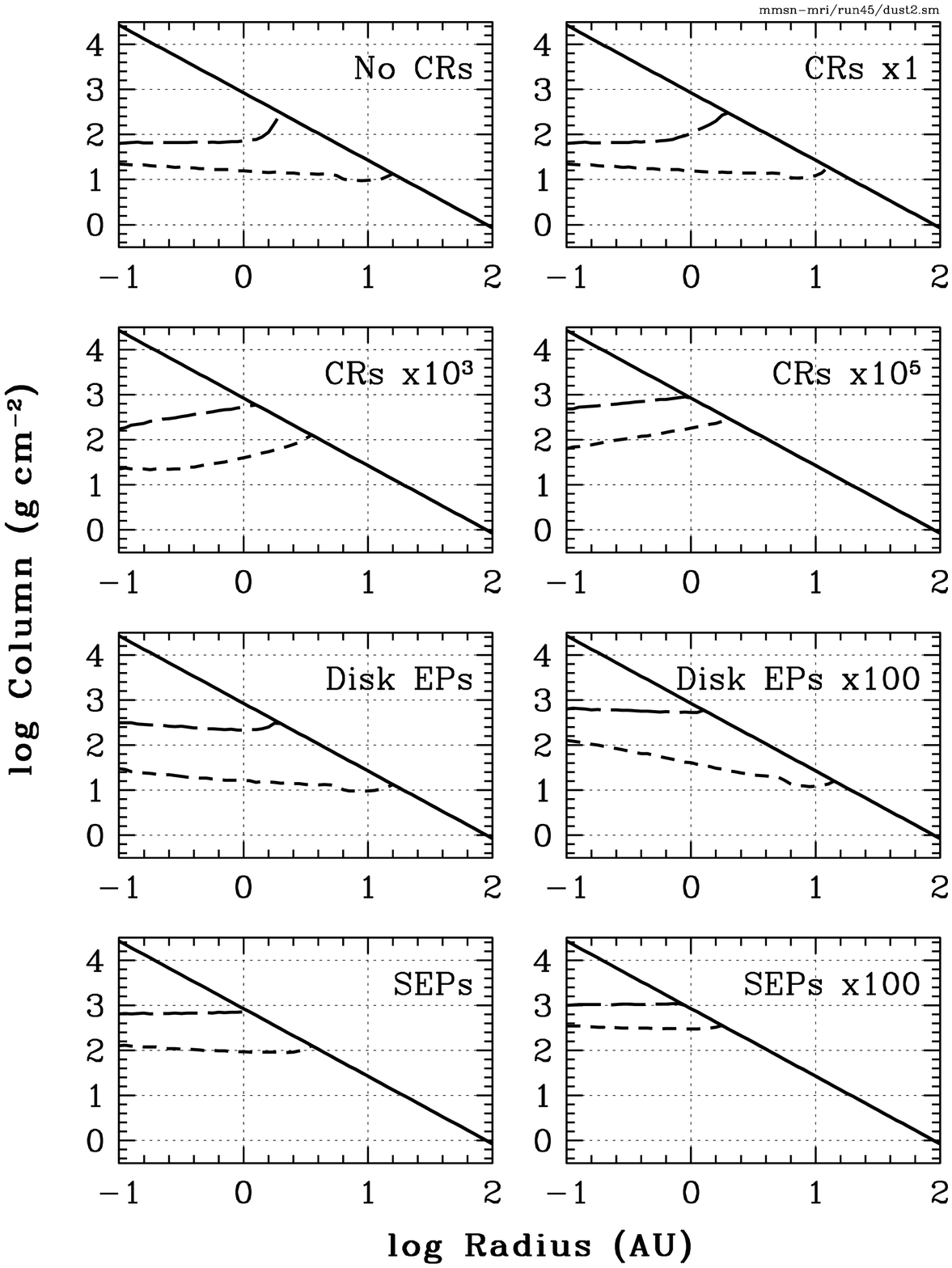}

  \figcaption{\sf Column depth to the top of the undead zone (short
    dashes) and dead zone (long dashes) in the minimum-mass protosolar
    disk with dust grains and short-lived rather than long-lived
    radionuclides.  The eight ionization scenarios are otherwise
    identical to figure~\ref{fig:dust}.
    \label{fig:dust2}}
\end{figure}

\begin{figure}[tb!]
  \epsscale{0.6}
  \plotone{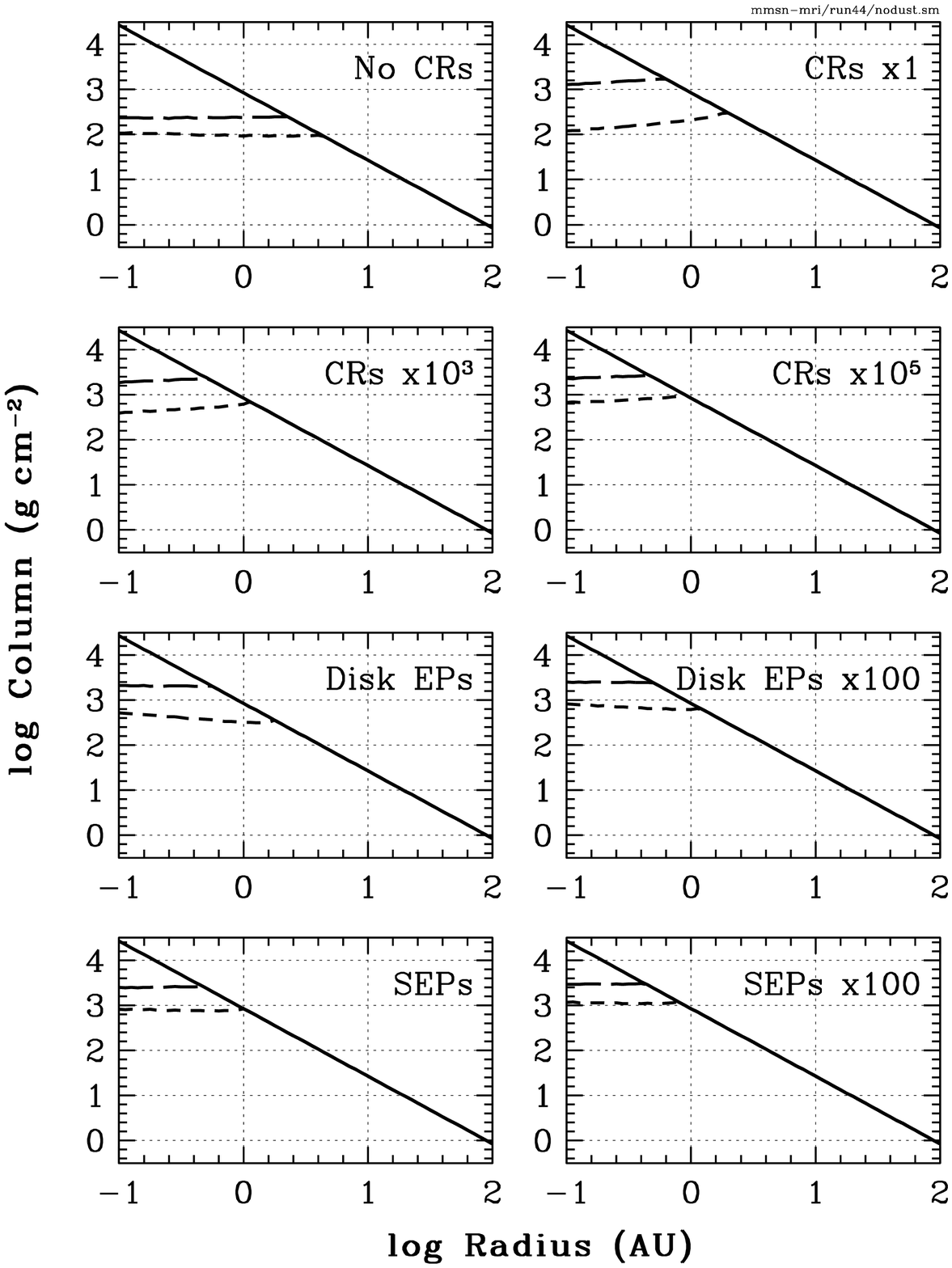}

  \figcaption{\sf Column depth to the top of the undead zone (short
    dashes) and dead zone (long dashes) in the minimum-mass protosolar
    disk without dust.  No radionuclide decay is included.  The eight
    ionization scenarios are otherwise identical to
    figure~\ref{fig:dust}.
    \label{fig:nodust}}
\end{figure}

\begin{figure}[tb!]
  \epsscale{0.6}
  \plotone{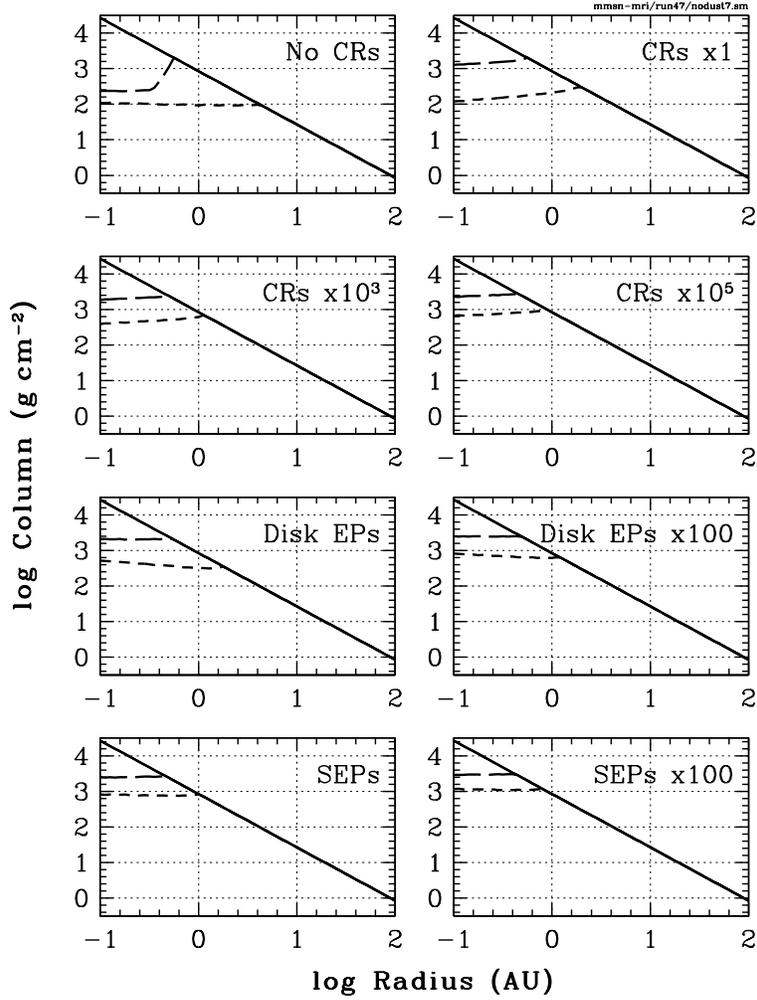}

  \figcaption{\sf Column depth to the top of the undead zone (short
    dashes) and dead zone (long dashes) in the minimum-mass protosolar
    disk without dust, assuming that 0.01\% of the long-lived
    radioactive $^{40}$K remains in the gas phase.  The eight
    ionization scenarios are otherwise identical to
    figure~\ref{fig:dust}.
    \label{fig:nodust2}}
\end{figure}

\begin{figure}[tb!]
  \epsscale{0.6}
  \plotone{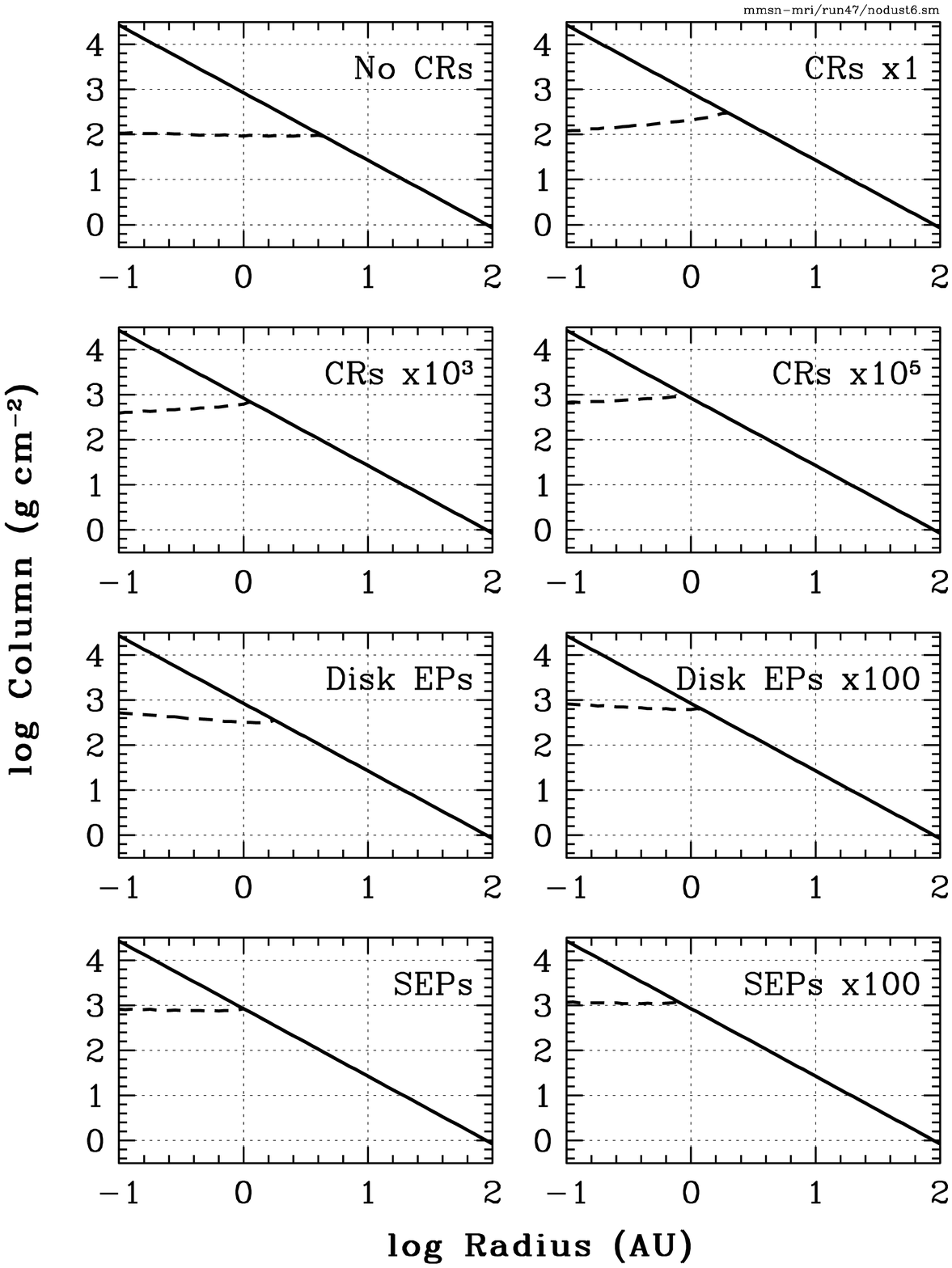}

  \figcaption{\sf Column depth to the top of the undead zone (short
    dashes) in the minimum-mass protosolar disk without dust, assuming
    that 0.01\% of the short-lived radioactive $^{26}$Al is found in
    the gas phase.  The eight ionization scenarios are otherwise
    identical to figure~\ref{fig:dust}.  The dead zone is eliminated
    under these conditions, but the undead zone remains.
    \label{fig:nodust3}}
\end{figure}

The dependence on the radionuclide ionization rates and on the
presence of grains is shown in the remaining figures.  Short-lived
radionuclides replace the long-lived radionuclides in
figure~\ref{fig:dust2}.  The dust is removed for
figures~\ref{fig:nodust} through~\ref{fig:nodust3}, which include no
radionuclides (figure~\ref{fig:nodust}), long-lived radionuclides with
an abundance reduced by a factor~$10^4$ below the dusty cases
(figure~\ref{fig:nodust2}) and short-lived radionuclides with the same
abundance reduction factor (figure~\ref{fig:nodust3}).

The results with and without dust are quite different.  With dust, the
active zone fills the top 10 to 20 g~cm$^{-2}$ if the external
ionization comes from stellar X-rays (figure~\ref{fig:dust}).  The
active zone is little changed by ordinary cosmic rays or by energetic
particles from the disk corona, since these are at most comparable to
the X-rays near the bottom of the active zone.  The active column
increases to 20 to 100 g~cm$^{-2}$ if the cosmic ray flux is boosted a
thousand times, and to 100 g~cm$^{-2}$ if the disk is exposed to the
protosolar energetic particles inferred from X-ray flares in young
stars.  The column of material lying in the undead zone is three to
ten times that in the active zone, while any remaining column is dead.
Including short-lived radionuclides moves the dead zone boundary
slightly deeper inside the disk while having little effect on the
outer boundary of the undead zone (figure~\ref{fig:dust2}).

With no dust, the active layer fills the outermost 100 g~cm$^{-2}$ if
the ionization is due to stellar X-rays, increasing to 100-200
g~cm$^{-2}$ if ordinary cosmic rays are added, and reaching 400-800
g~cm$^{-2}$ under an enhanced flux of cosmic rays, disk energetic
particles, or protosolar energetic particles
(figure~\ref{fig:nodust}).  The undead column is two to ten times that
in the active layer.

Including a small amount of radioactivity leaves the active layer
unchanged because the radionuclide ionization rate is much less than
that due to X-rays at the active layer boundary.  The extra ionization
can be important however in the regions shielded from the energetic
protons and photons, yielding a bigger undead zone at the expense of
the dead zone.  The differences are slight with 0.01\% of the
long-lived radionuclides present in the gas
(figure~\ref{fig:nodust2}).  By contrast, the whole dead zone becomes
undead if 0.01\% of the short-lived radionuclides is included
(figure~\ref{fig:nodust3}).  Reducing the fraction to 0.001\% leads to
the appearance of a small dead zone just outside 0.1~AU.  The
threshold internal ionization rate for converting the entire dead zone
into an undead zone is thus about $3\times 10^{-5}\zeta_{SR} \approx
10^{-23}$~s$^{-1}$.

Cosmic rays from a nearby supernova can increase the active column by
up to a factor eight over the interstellar cosmic ray case if the base
of the active layer is defined solely by the cosmic ray ionization
rate \citep{fa06}.  When the stellar X-rays and the rise in
recombination rate with depth are included, the higher cosmic ray flux
increases the active column at 1~AU only from 16 to 41 g~cm$^{-2}$
with dust (figure~\ref{fig:dust}) and from 210 to 620 without
(figure~\ref{fig:nodust}), factors of about three.  The thousand-fold
enhancement in the cosmic ray flux has less effect than removing the
dust grains.

\section{CONCLUSIONS\label{sec:conclusions}}

We calculated the locations of the undead and dead zones in the
minimum-mass protosolar disk, under ionization scenarios including
stellar X-rays, long- or short-lived radionuclide decay, and energetic
protons arriving from the general interstellar medium, from a nearby
supernova explosion, from the disk corona, or from the corona of the
young star.

The disk contains an undead zone in all the scenarios.  All have a
dead zone except those with the dust removed and a small fraction of
radioactive $^{26}$Al remaining in the gas.  Lacking short-lived
radionuclides, even the strongest external ionization process
considered, the stellar energetic particles, fails to eliminate the
dead zone.  The thickness of the magnetically coupled layers can
increase over the local chemical equilibrium values we have computed,
however, if ionized material mixes to the midplane before recombining.
Slow recombination requires that dust is largely absent
\citep{is05,in06b,ts07,in08}.

Placing these results in the context of existing work, we see that
protostellar disks with no dead zone can be constructed by
\begin{enumerate}
\item Raising the temperature of the high-column material sufficiently
  for thermal ionization.  The temperatures required are likely to be
  reached through either stellar illumination or accretion heating
  only within a few tenths of an AU of the star, at average mass flow
  rates.  At the highest rates inferred for episodic accretion,
  thermal ionization of the interior may extend to a few AU.  The
  central region could alternate between a cold state with a dead zone
  and a low accretion rate, and a hot, thermally-ionized state with a
  higher accretion rate \citep{al01,zh09}.
\item Reducing the surface density below the minimum-mass protosolar
  model \citep{ft02}.  Forming planets under these conditions may be
  problematic.  Lower surface densities could occur however during the
  dispersal of the disk, for example as X-ray ionized material is
  eroded from the inner edge \citep{cm07}.
\item Removing the small dust grains while retaining a fraction of the
  radionuclides in the gas.  The removal could occur through settling
  in the dead zone \citep{c07} or incorporation into planetesimals.
  Making grains unimportant compared with gas-phase recombination
  requires reducing the dust abundance by several orders of magnitude
  \citep{sm00}, a difficult feat during planet formation, when solid
  bodies often collide fast enough to shatter and produce secondary
  particles \citep[e.g.][]{ws93}.
\end{enumerate}
Each of these three possibilities appears feasible only in limited
circumstances.  A fourth option involves ionization by the electrons
carrying currents in the weakly-ionized disk interior \citep{is05}.
Further study of the microphysics of this process is desirable.  We
conclude that the dead and undead zones are robust consequences of
cold, dusty gas without strong internal sources of ionization, having
mass columns exceeding 1000 g~cm$^{-2}$.  Weaker turbulence at the
midplane and reduced or fluctuating mass flows are likely under these
conditions, and should be considered further in planet formation
models.

\acknowledgments

We thank Eliot Quataert and the staff of the Theoretical Astrophysics
Center at the University of California, Berkeley for their hospitality
and comments, and Eric Feigelson, Hugh Hudson and Mark Wiedenbeck for
useful discussions.  The work was carried out in part at the Jet
Propulsion Laboratory, California Institute of Technology, with
support from the NASA Solar Systems Origins Program.  Copyright 2009.
All rights reserved.


\end{document}